\documentclass[12pt,preprint]{aastex}
\usepackage{epsfig}
\begin{document}
\voffset-1cm
\newcommand{\gsim}{\hbox{\rlap{$^>$}$_\sim$}}
\newcommand{\lsim}{\hbox{\rlap{$^<$}$_\sim$}}

\title{What did we learn from the extremely bright gamma ray bursts
       990123 and 080319B?}

\author{Shlomo Dado\altaffilmark{1} and Arnon Dar\altaffilmark{2}}

\altaffiltext{1}{dado@phep3.technion.ac.il\\
Physics Department, Technion, Haifa 32000,
Israel}
\altaffiltext{2}{arnon@physics.technion.ac.il\\
Physics Department, Technion, Haifa 32000}

\begin{abstract}

The two brightest and so far the best studied gamma ray bursts (GRBs), 
990123 and 080319B, were ordinary, highly collimated GRBs produced in a 
core collapse supernova explosion within a high-density wind environment 
and observed from a very near-axis viewing angle. Inverse Compton 
scattering (ICS) and synchrotron radiation (SR), the two dominant 
radiation mechanisms in the cannonball (CB) model of GRBs, together with 
the burst environment, provide a very simple and sufficiently accurate 
description of the multiwavelength lightcurves of their prompt and 
afterglow emissions.

\end{abstract}


\section{Introduction}

Two models have been used extensively to analyze Gamma ray bursts (GRBs)  
and their afterglows (AGs), the fireball (FB) model (for recent reviews 
see, e.g., M\'{e}sz\'{a}ros~2002, 2006; Zhang~2007) and the cannonball 
(CB) model (Dar \& De R\'ujula~2004, hereafter DD2004; Dado \& Dar 
2008a,b, hereafter DD2008a,b, and references therein). Despite their 
similar names, the two models are (or were) entirely different (e.g., 
Piran~1999), hence only one of them, if either, may provide a faithful 
physical description of GRBs. Until recently, the FB model has been widely 
accepted as that model. However, the rich data on GRBs accumulated from 
space based observations, in particular from the Swift satellite, 
complemented by early time data from ground based robotic telescopes and 
late-time follow-up observations with larger telescopes, have challenged 
this prevailing view. Kumar et al.~(2007) concluded that the prompt 
$\gamma$-ray emission cannot be produced in shocks, internal or external. 
Zhang, Liang and Zhang~(2007) found that the fast decay and rapid spectral 
softening ending the prompt gamma-ray and X-ray emission cannot be 
explained simultaneously by high latitude emission. The X-ray and optical 
afterglows (AGs) of the Swift GRBs were found to be chromatic at early 
time (Covino et al.~2006) and to have chromatic breaks (Panaitescu et 
al.~2006) which differ significantly from the jet breaks expected in the 
collimated fireball model of AGs. Burrows and Racusin~(2007) examined the 
XRT light curves of the first $\sim\! 150$ Swift GRBs and reported that 
the expected jet breaks are extremely rare. In particular, Liang et 
al.~(2008) have analyzed the Swift X-ray data for the 179 GRBs detected 
between January 2005 and January 2007 and the optical AGs of 57 pre- and 
post-Swift GRBs. They found that not a single burst satisfies all the 
criteria of a jet break.

In spite of the above, not all authors are so critical and they believe that 
the GRB data require only some modifications of the standard FB model in 
order to accommodate the results (e.g.,~Panaitescu~2006; Dai et 
al.~2007; Sato et al.~2007; Kumar and Panaitescu~2008). Other authors seem 
to ignore the failures of the FB model and continue to interpret the 
observations with the FB model hypotheses (`internal and external shocks', 
`colliding conical shells', `forward and reverse shocks', `continuous 
energy injection') and parametrize the data with freely adopted formulae 
(e.g., `segmented power laws') which were never derived explicitly from 
underlying physical assumptions (for recent examples, see, e.g., Bloom et 
al.~(21 authors)~2008; Racusin et al.~(93 authors)~2008).

The situation concerning the CB model is different. The predictions of the 
model were derived in fair approximations from its underlying assumptions. 
They were shown to describe correctly the main observed properties of GRBs 
and reproduce successfully the diverse broad band light curves of large 
representative sets of both long GRBs and XRFs (e.g., DD2004; DD2008a and 
references therein) and short hard bursts (DD2008b). In this paper, we 
demonstrate these for the GRBs 990123 and 080319B, the brightest GRBs 
observed so far that have the best sampled and the most accurately 
measured multiwavelegth light curves of the prompt and afterglow emission. 
We show that the two underlying radiation mechanisms of the CB model, 
inverse Compton scattering (ICS) and synchrotron radiation (SR), and the 
burst environment suffice within the CB model provide a very simple and 
successful description of the observed light curves of their prompt and 
afterglow emissions.

\section{GRBs 990123 and 080319B} 

GRB 990123, at redshift $z\! =\! 1.600$ (Kulkarni et al.~1999a; Andersen 
et al.~1999), with the highest known gamma ray energy under the assumption 
of 
isotropic emission, $E_{iso}\!\approx\!2\times10^{54}$ erg, was detected 
and localized by the Burst And Transient Source Experiment (BATSE) on 
board the Compton Gamma Ray Observatory (CGRO), which measured its light 
curves 
in the 20-50 keV, 50-100 keV, 100-300 keV and $>$300 keV BATSE channels 
(see e.g., Fig.~3), and at higher energies by the COMPTEL, OSSE and 
EGRET instruments (Briggs et al.~1999). It was also detected and localized 
by the Gamma Ray Burst Monitor (GRBM) aboard the BeppoSAX satellite 
(Maiorano et al.~2005), which measured its lightcurve and spectrum in the 
range 2-700 keV 
for the first 80 s before the burst entered Earth occulation. The BeppoSAX 
measurements were resumed after it came out of the Earth occulation and 
ended 2.49 days after burst because of diminishing statistics. GRB 990123 
was also the first GRB in which an optical emission was detected during 
the prompt $\gamma$/X-ray emission. The emission, which was detected by 
the Robotic Optical Transient Search Experiment (ROTSE) at Los Alamos 
National Laboratories, was triggered by BATSE 22 s after the onset of the 
burst, brightened and peaked at magnitude $V\!\sim\! 9$, about 50 s after 
the GRB onset, and decayed rapidly with time (Akerlof et al.~1999) which 
was followed in the $UVONIR$ bands with large ground based telescopes 
(Castro-Tirado et al.~1999; Galama et al.~1999; Kulkarni et al.~1999a; 
Fruchter et al.~1999; Holland et al.~2000 and references therein) and with 
the Hubble Space Telescope until it faded to a magnitude 
$V\!=\! 27.7\!\pm\!0.15$, two months after burst 
(Fruchter et al.~2000). The broad band 
$\gamma$-ray, X-ray, $UVO$ and $NIR$ lightcurves of GRB 990123 were 
reanalyzed recently within the synchrotron fireball (FB) model by 
Corsi et al.~(2005). Essentially they found that the 
spectral and temporal properties of the prompt optical emission are 
uncorrelated to the $\gamma$ and X-ray emission, implying different 
physical origins, that the optical and X-ray afterglow lightcurves are 
chromatic contrary to expectations and that their spectral and temporal 
power-law decays do not satisfy the correlations expected in the FB model.

GRB 080319B at redshift $z\!=\!0.937$ was simultaneously detected by the 
Swift-Burst Alert Telescope (BAT) and the Konus gamma-ray detector aboard 
the Wind satellite (Racusin et al.~2008; Golenetskii et al.~2008). The 
location of GRB 080319B was fortuitously only 10$^o$ away from GRB 
080319A, 
which was detected by Swift less than 30 minutes earlier, and allowed 
several wide field telescopes to detect the optical emission of GRB 
080319B instantly. It peaked 26 s after the Swift trigger at magnitude 
$V\!=\!5.3$ (Racusin et al.~2008; Wozniak et al.~2008) visible to the 
naked eye. The extreme 
brightness of the burst and its gamma-ray, X-ray and $UVOIR$ afterglows 
led to a flurry of follow-up observations with a variety of space-  
and 
ground-based telescopes, which were summarized in Bloom et al.~2008,
Racusin 
et al.~2008 and Wozniak et al.~2008. GRB 080319B was quite similar to GRB 
990123 in many respects. Its isotropic equivalent gamma-ray energy release 
was $E_{iso}\!\approx\! 1.3\times 10^{54}$ erg, similar to that of GRB 
990123. Like in GRB 990123, the fast spectral variation of its hard X-ray 
and gamma ray emission was well parametrized with an exponentially 
cut-off power-law with a cut-off energy that was strongly correlated with 
the peak 
structure of the lightcurve and a low-energy photon spectral index, 
$\Gamma\!\approx\!1$, which changed abruptly into $\Gamma\!\approx\!2.1$ 
after the fast decay phase of the prompt emission (e.g., Fig.~10). The 
optical and gamma-ray light curves during the explosion were not 
correlated (see, e.g., Fig.~1 in Racusin et al.~2008): The onset of the 
optical emission lagged behind the gamma ray emission by several seconds 
and decayed more slowly at the end of the prompt emission. The typical 
time scales of their temporal variability were entirely different. The 
extremely bright optical emission could not be reconciled with a single 
emission mechanism - extrapolating the gamma-ray spectrum to the optical 
band underestimates the optical flux by more than 4 orders of magnitude. 
Their spectra were also quite different. Contrary to expectations, the 
X-ray and $UVO$ afterglow light curves were also chromatic, with no `jet 
breaks' and with spectral and temporal power-law decays which did not 
satisfy the correlations expected in the FB model (see, however, Bloom et 
al.~2008 (21 authors); Racusin et al.~2008 (93 authors!); Wozniak et 
al~2008; Kumar and Panaitescu~2008 for attempts to reconcile the 
observations with the FB model).

\section{The CB model}

In the CB model (e.g.,~DD2004 and references therein)  {\it long-duration} 
GRBs and their AGs are produced by bipolar jets of highly relativistic 
plasmoids of ordinary matter (Shaviv \& Dar~1995, Dar~1998) ejected in 
core-collapse supernova (SN) explosions (Dar \& Plaga~1999). An accretion 
disk or a torus is hypothesized to be produced around the newly formed 
compact object, either by stellar material originally close to the surface 
of the imploding core and left behind by the explosion-generating outgoing 
shock, or by more distant stellar matter falling back after its passage 
(De R\'ujula~1987). As observed in microquasars (e.g.,~Mirabel \& 
Rodriguez~1999, Rodriguez \& Mirabel~1999 and references therein), each 
time part of the accretion disk falls abruptly onto the compact object, 
two jets of cannonballs (CBs) made of {\it ordinary-matter plasma} are 
emitted with large bulk-motion Lorentz factors in opposite directions 
along the rotation axis, wherefrom matter has already fallen back onto the 
compact object due to lack of rotational support. The prompt $\gamma$-ray 
and X-ray emission is dominated by inverse Compton scattering (ICS) of 
photons of the SN glory - light scattered/emitted by the pre-supernova 
wind blown from the progenitor star. The CBs' electrons Compton up-scatter 
the glory photons to $\gamma$-ray energies and collimate them into a 
narrow beam along the CBs' directions of motion.

A second mechanism besides ICS that generates radiation by a CB is 
synchrotron radiation (SR). The CBs which initially expand in their rest 
frame with the speed of sound in a relativistic plasma, collide and merge 
in a short time into a few leading CBs. The jet of CBs encounters matter 
in its voyage through the wind/ejecta blown by the progenitor star and 
then through the interstellar medium (ISM), which is ionized by the beamed 
radiation of the very same CBs. The ISM ions continuously impinging on a 
CB generate within it a turbulent magnetic field, which is assumed to be 
in approximate energy equipartition with their energy. In this field the 
Fermi accelerated CB and ISM intercepted electrons emit synchrotron 
radiation. The initial expansion of a CB produces a rapidly rising 
lightcurve which stops rising when the expansion slows down. When the CB 
expansion stops, its emission turns into a decline which traces the 
circumburst density of the pre-supernova wind/ejecta blown by the 
progenitor star into the roughly constant ISM density. Only when the CB 
has swept a mass comparable to its rest mass, does the continuous 
collision with the medium begins to decelerate it effectively, which 
results in a gradual steepening (break) of its SR afterglow lightcurve 
into an asymptotic power-law decay.

\section{Correlations between the prompt emission observables}

Straightforward consequences of the CB model are simple correlations 
among various properties of the prompt emission pulses
(Dar \& De R\'ujula 2000, hereafter DD2000). For instance, 
the relativistic boosting and beaming of the glory photons by a CB yield 
the relations (DD2000; DD2004), 
\begin{equation} 
      E_{iso}\propto \delta^3\,;~~~~
     (1+z)^2\, L_p\propto \delta^4\,;~ ~~~~
     (1+ z)\, E_p \propto \gamma\, \delta\,, 
\label{correlation} 
\end{equation}
where $E_{iso}$ is the isotropic equivalent gamma ray energy, $L_p$ is the 
peak isotropic equivalent luminosity, $E_p$ is the photon energy at peak 
energy flux density of an ICS pulse, $\gamma$ is the bulk motion Lorentz 
factor of a CB, and $\delta\! =\! 1/\gamma\, (1\!-\!\beta\, cos\theta)$ is 
its Doppler factor with $\theta$ being the angle between the line of sight 
to the CB and its direction of motion. For $\gamma^2 \gg 1$ and $\theta^2 
\ll 1$, $\delta \approx 2\, \gamma/(1\!+\!\gamma^2\, \theta^2)$ to an 
excellent approximation. The strong dependence of observables such as 
$E_{iso}\,,$ $L_p$ and $E_p$ on $\gamma$ and $\delta$ and the 
narrow distribution of $\theta$ around $1/\gamma$  result in 
correlations among them. The observed correlations between $(1 + z)\, E_p$ 
and $E_{iso}$ and between $(1 + z)\, E_p$ and $(1+z)^2\, L_p$ in GRBs with 
known redshift are shown in Figs.~1a,1b. The correlations predicted by 
the CB model (e.g. DD2000; DD2004; Dado, Dar \& De R\'ujula 2007,
hereafter DDD2007), which are 
indicated by the thick lines in these figures, well represent the observed 
correlations (e.g., Amati 2002,2006) and
are also well satisfed by GRBs 990123 and 080319B. The 
locations of GRBs 990123 and 080319B in these correlation plots and in the 
other well established correlations in GRBs (see, e.g., DDD2007) suggest 
that GRBs 990123 and 080319B were probably ordinary GRBs viewed very near 
axis, $\theta^2\!\ll\! 1/\gamma_0^2\, :$ The most probable viewing angle 
of GRBs, $\theta\approx 1/\gamma$, yields $\delta\approx \gamma$, while 
for small viewing angles, $\theta^2 \!\ll\! 1/\gamma^2$ and then 
$\delta\!\approx\! 2 \,\gamma$, yielding a rest frame $E_p$, $E_{iso}$ and 
$L_p$, which are 2 times larger, 8 times more energetic and 16 times 
brighter, respectively, than their mean values observed in long GRBs with 
the most probable $\gamma\, \theta\approx 1$. The inferred large Lorentz 
and Doppler factors of GRBs 990123 and 080319B, coupled with a 
high-density wind/ejecta blown by their progenitor stars before the 
supernova explosion, explains their initially extreme optical 
brightness and the lack of `jet breaks' in their afterglows 
(Dado, Dar \& De R\'ujula 2008b, hereafter DDD2008b).

\section{The spectrum of ICS pulses}

The predicted time-dependent spectrum of a GRB pulse 
is given by (DD2004):
\begin{equation}
E\, {dN_\gamma\over dE} \sim \left({E\over E_p(t)}\right)^{-\beta_g}\,
 e^{-E/E_p(t)}+ b\,(1-e^{-E/E_p(t)})\, \left({E \over
E_p(t)}\right)^{-p/2}\,.
\label{GRBspec}
\end{equation}
The first term in Eq.~{\ref{GRBspec} with $\beta_g\!\sim\! 0$, is the
result of inverse Compton scattering of 
glory photons with a thin bremsstrahlung spectrum,
\begin{equation}
\epsilon\, {dn_\gamma \over d\epsilon} \approx n_\gamma(r)\,
\left({\epsilon \over T_g }\right)^{-\beta_g}\, e^{-\epsilon/T_g},
\label{thinbrem}
\end{equation} 
by the bulk of the CB's electrons, which are comoving with it.
The second term in Eq.~(\ref{GRBspec}) is induced by
a very small fraction of
`knocked on' and Fermi accelerated electrons, whose initial spectrum
(before Compton and synchrotron cooling) is $dN_e/dE\propto E^{-p}$,
with $p\approx 2.2$.
The effective temperature $T_g(r)$ of the glory 
that decreases with distance yields (DD2008b)
\begin{eqnarray}
E_p(t)&\approx&  E_p(0)\, {t_p^2 \over t^2+t_p^2}\,,
\nonumber\\
E_p(0)&\approx & {\gamma_0\, \delta_0 \over 1+z}\, T_g(0),
\label{PeakE}
\end{eqnarray}
where typically, $T_g(0)\!\sim\!1$ eV, and  
$t_p$ is the time when the ICS contribution 
to $E\, d^2N_\gamma/ dE\, dt\,, $ reaches its peak value. 

For $\beta_g=0$, the peak energy of the time integrated spectrum 
of the ICS contribution, 
$E_p\!=\!max\,  E^2\int (d^2N_\gamma/ dE\, dt)\, dt\,,$ 
satisfies $E_p\!=\!E_p(t_p)$ (see Appendix I).

For $b={\cal{O}}(1)$,
the energy spectrum predicted by the CB model, Eq.~(\ref{GRBspec}),
bears a striking resemblance
to the Band function (Band et al.~1993) traditionally used to model the
energy spectra of GRBs,
but GRBs where the spectral measurements
extended over a much wider energy range than that of BATSE and Swift/BAT,
are better fit by Eq.~\ref{GRBspec} (e.g., Wigger et al.~2008).
Moreover, the spectral evolution during the prompt emission pulses
is well described by Eqs.~(\ref{GRBspec}) and (\ref{PeakE})
(Dado, Dar and De Ru\'jula 2008a, hereafter DDD2008a; DD2008a).

\section{The lightcurve of an ICS pulses}

An ICS pulse has an approximate lightcurve (DD2008a),
\begin{equation}
 E\,{d^2N_\gamma \over dt\, dE} \propto {\Delta t^2\, t^2  \over
(t^2+\Delta t^2)^2}\,E\,{dN_\gamma\over dE}\, ,
\label{ICSPulse}
\end{equation}
where $E\,{dN_\gamma/ dE}$ is given by Eq.~(\ref{GRBspec}).
At the relatively low X-ray energies covered by Swift and,
more so, at smaller ones, the first term on the RHS of
Eq.~(\ref{GRBspec}) usually dominates  $E\, dN_\gamma/dE$. Consequently,
the lightcurve generated by a  sum of ICS pulses
at a luminosity distance $D_L$ is generally well approximated by:
\begin{equation}
E\,{d^2N_\gamma \over dt\, dE} \approx \Sigma_i\,
A_i\Theta[t\!-\!t_i]\,{\Delta t_i^2\,(t\!-\!t_i)^2 \over
((t\!-\!t_i)^2\!+\!\Delta t_i^2)^2}\, e^{-E/E_{p,i}[t\!-\!t_i]}\, ,
\label{ICSlc}
\end{equation}
where the index `i' denotes the i-th pulse produced by a
CB launched at an observer time $t\!=\!t_i$, 
$E_{p,i}[t\!-\!t_i]$ is given by Eq.~(\ref{PeakE}) with $t$ replaced
by $t\!-\!t_i$ and $A_i$ is a constant that depends on the 
radius of the CB, its Lorentz and Doppler factors, the density 
of the glory light and the redshift and distance of the GRB
(DD2008a).
Thus, in the CB model, each ICS pulse in the GRB light curve
is effectively described by four parameters, $t_i,\, A_i,\,
\Delta t_i$  and $E_{p,i}(0)$,
which are a-priori unknown and thus are best fit to reproduce its observed 
light curve.

Setting $t_i=0$, $E_p(t)$  has the approximate form
$E_p(t)\!\approx\! E_p(0)\, t_p^2/(t_p^2\!+\!t^2).$
Such an evolution
has been observed in the time-resolved spectrum of well isolated pulses
(see, for instance, the insert in Fig.~8 of Mangano et al.~2007),
until the ICS emission is overtaken by the broad band
synchrotron emission from the swept-in ISM electrons. Hence,
the temporal behaviour of the  separate ICS peaks at $E<E_p$
is given by:
\begin{equation}
E\, {d^2N_\gamma\over dt\,dE}(E,t)
\propto {t^2/\Delta t^2  \over(1+t^2/\Delta t^2)^2}\,
e^{-E\, (t^2+t_p^2)/ E_p(0)\,t_p^2}\approx e^{-E/E_p(0)}\, F(E\,t^2)
\approx F(E\,t^2),
\label{law}
\end{equation}
to which we shall refer as the `$E\,t^2$~{\it law'}.
A simple consequence of this law is that unabsorbed ICS peaks 
for $E<E_p$ have
approximately identical shape at different energies when plotted as
a function of $E\, t^2$.

A few other trivial but important consequences of Eq.~(\ref{law})
for unabsorbed GRB peaks at $E \lsim E_p$ are the following:
\begin{itemize}

\item{}
The peak time of a pulse is at,
\begin{equation}
t_p=t_i\!+\!\Delta t_i\,.
\label{peaktime}
\end{equation}

\item{}
The full width at half maximum (FWHM) of a  pulse is,
\begin{equation}
{\rm FWHM}\!\approx\! 2\, \Delta t_i,
\label{fwhm}
\end{equation}
and it extends from $t\!\approx\!t_i\!+\! 0.41\, \Delta t_i $ to
$t\!\approx\!t_i\!+\! 2.41\,\Delta t_i)$.

\item{}
The rise time (RT) from half peak
value to peak value satisfies,
\begin{equation}
{\rm RT\approx 0.30\,FWHM},
\label{ratio}
\end{equation}
independent of energy. It agrees with the empirical
relation that was inferred by Kocevski et al.~(2003)
from BATSE bright GRBs,
${\rm RT\!\approx\! (0.32\!\pm\!
0.06)\, FWHM}$.

\item{}
The FWHM increases
with decreasing energy approximately like a power-law,
\begin{equation}
{\rm FWHM}(E)\sim E^{-0.5}\, .
\label{widthrelation}
\end{equation}
This relation is consistent with the empirical relation,
${\rm FWHM}(E) \propto E^{\!-\!0.42\!\pm\! 0.06}$,
satisfied by BATSE GRBs (Fenimore et al.~2003).

\item{}
The onset-time, $t_i$, of a pulse is simultaneous at all energies.
But the peak times $t_p$ at different energies differ and the
lower-energy ones `lag'
behind the higher-energy ones:
\begin{equation}
t_p-t_i \propto E^{-0.5}\, .
\label{lagtime}
\end{equation}

\item{}
The time averaged value of $E_p(t)$ for GRB peaks, which follows from
Eq.~(\ref{PeakE}), satisfies:
  \begin{equation}
E_p= E_p(0)/2=E_p(t_p)\, .
\label{Epeak}
\end{equation}
\end{itemize}

\section{The emission of synchrotron radiation}
\label{Synchrotron}

The ISM ions continuously impinging on
a CB generate
within it turbulent magnetic fields, which are assumed to be
in approximate energy equipartition with their energy,
$B\!\approx\! \sqrt{4\,\pi\, n\, m_p\, c^2}\, \gamma$.
In this field, the intercepted
electrons emit synchrotron radiation. The SR, isotropic in the CB's
rest frame, has a characteristic frequency, $\nu_b(t)$,
the typical frequency radiated by the
electrons that enter a CB at time $t$ with a relative Lorentz
factor $\gamma(t)$. In the observer's frame:
\begin{equation}
\nu_b(t)\simeq  {\nu_0 \over 1+z}\,
{[\gamma(t)]^3\, \delta(t)\over 10^{12}}\,
\left[{n\over 10^{-2}\;\rm cm^3}\right]^{1/2}
{\rm Hz},
\label{nub}
\end{equation}
where $\nu_0\!\simeq\! 3.85\times 10^{16}\, {\rm Hz}\!\simeq\! 160/h$ eV.
The spectral energy density of the SR
from a single CB at a luminosity distance $D_L$  is given by (DDD2003a):
\begin{equation}
F_\nu \simeq {\eta\,  \pi\, R^2\,n\, m_e\, c^3\,
\gamma(t)^2\, \delta(t)^4\, A(\nu,t)\,
\over 4\,\pi\, D_L^2\,\nu_b(t)}\;{p-2\over p-1}\;
\left[{\nu\over\nu_b(t)}\right]^{-1/2}\,
\left[1 + {\nu\over\nu_b(t)}\right]^{-(p-1)/2}\,,
\label{Fnu}
\end{equation}
where $p\!\sim\! 2.2$ is the typical spectral index
of the Fermi accelerated
electrons, $\eta\!\approx\!1$ is the fraction of the impinging ISM
electron energy that is synchrotron re-radiated by the CB, and $A(\nu, t)$
is the attenuation of photons of observed frequency $\nu$ along the line
of sight through the CB, the host galaxy (HG), the intergalactic medium
(IGM) and the Milky Way (MW):
\begin{equation}
A(\nu, t) = {\rm
exp[-\tau_\nu(CB)\!-\!\tau_\nu(HG)\!-\!\tau_\nu(IGM)\!-\!\tau_\nu(MW)].}
\label{attenuation}
\end{equation}
The opacity $\tau_\nu\rm (CB)$ at very early times, during the
fast-expansion phase of the CB, may strongly depend on time and frequency.
The opacity of the circumburst medium [$\tau_\nu\rm (HG)$ at early times]
is affected by the GRB and could also be $t$- and $\nu$-dependent.  The
opacities $\tau_\nu\rm (HG)$ and $\tau_\nu\rm (IGM)$ should be functions
of $t$ and $\nu$, for the line of sight to the CBs varies during the AG
observations, due to the hyperluminal motion of CBs.

\subsection{Early-time SR}

The initial rapid expansion of a CB slows down as it propagates
through the wind and scatters its particles
(DDD2002, DD2004).  This expansion is roughly  described by
$R^2\approx R_{cb}^2\, t^2/(t^2\!+\!t_{exp}^2)$, where $R_{cb}$
is the asymptotic radius of the CB.
The effective deceleration of a CB begins only when it has swept 
a mass comparable to its rest mass (see next subsection).   
Until that time  both $\gamma$
and $\delta$ stay put at their initial values
$\gamma_0$ and $\delta_0$ and
Eq.~(\ref{Fnu})
yields an early-time SR light curve,
$F_\nu\! \propto\! e^{-\tau_{_W}}\, R^2\, n^{(1+\beta)/2}\,\nu^{-\beta}.$
where $\beta(t)\!=\!0.5$ for $\nu\!\ll\!\nu_b(t)$ and 
$\beta(t)\!=\!p/2$
for $\nu\!\gg\!\nu_b(t)$. 
Since $r\!\propto\!t$, a CB ejected into
a windy density profile, $n\!\propto\!1/r^2$,
created by the mass ejection from the
progenitor star prior to its SN explosion, emits SR
with an early-time light curve,
\begin{equation}
F_\nu \propto  {e^{-a/t}\,
t^{1-\beta} \over t^2+t_{exp}^2}\, \nu^{-\beta}\, .
\label{SRP}
\end{equation}
For a jet of CBs ejected at times $t_i$, the early SR lightcurve becomes 
the sum of such contributions from the individual CBs with 
the time  $t$ replaced by, $t\!-\!t_i$, the times after their ejection.

\subsection{SR during the CB's coasting phase}
As it plows through the ionized ISM, a CB
gathers and scatters its constituent ions, mainly protons.(DD2004). The 
scattered and re-emitted
protons exert an inward pressure on the CB, countering its expansion.
In the approximation of isotropic re-emission in the CB's
rest frame and a constant ISM density $n\!\sim\!n_e\!\sim\!n_p$,
one finds that, within minutes of observer's time $t$, a typical CB
of baryon number $N_b\!\approx 10^{50}$ reaches an approximately constant
`coasting' asymptotic radius $R\!\sim\!10^{14}$ cm, before it finally
stops and blows up, after a journey of months of observer's
time. During the coasting phase, and in a constant density ISM,
$\gamma(t)$  obeys (Dado et al.~2006):
\begin{equation}
({\gamma_0/ \gamma})^4+
2\,\theta^2\,\gamma_0^2\,(\gamma_0/\gamma)^2
=1\!+\!2\!\,\theta^2\,\gamma_0^2\!+\!t/t_0\,,
\label{decel}
\end{equation}
the solution of which is, 
\begin{equation}
\gamma(t) = {\gamma_0\over [\sqrt{(1+\theta^2\,\gamma_0^2)^2 +t/t_0}
          - \theta^2\,\gamma_0^2]^{1/2}}\,,
\label{goft}
\end{equation}
with
\begin{equation}
t_0={(1\!+\!z)\, N_{_{\rm B}}\over 8\,c\, n\,\pi\, R^2\,
\gamma_0^3}\,.
\label{break}
\end{equation}
The deceleration law is for the
case in which the ISM particles re-emitted fast by the
CB are a small fraction of the flux of the intercepted ones.
As can be seen from Eq.~(\ref{decel}), $\gamma$  and $\delta$
change little as long as $t\!\ll\! t_b\!=\![1\!+\gamma_0^2\,
\theta^2]^2\,t_0\, .$
In terms of typical CB-model values of $\gamma_0$,
$R$, $N_{_{\rm B}}$ and $n$,
\begin{equation}
t_b= (1300\,{\rm s})\, [1+\gamma_0^2\, \theta^2]^2\,(1+z)
\left[{\gamma_0\over 10^3}\right]^{-3}\,
\left[{n\over 10^{-2}\, {\rm cm}^{-3}}\right]^{-1}
\left[{R\over 10^{14}\,{\rm cm}}\right]^{-2}
\left[{N_{_{\rm B}}\over 10^{50}}\right] \! .
\label{tbreak}
\end{equation}
For $t\!\gg\!t_b$, $\gamma$  and $\delta$ decrease like $t^{-1/4}\,.$ 
The transition $\gamma(t)\!\sim\! \gamma_0\! \rightarrow\!\gamma\!\sim\!
\gamma_0\,(t/t_0)^{-1/4}$
induces a bend (the so called `jet  break') 
in the synchrotron AG from a plateau to an asymptotic power-law
decay,
\begin{equation}
F_\nu \propto t^{-p/2-1/2}\,\nu^{-p/2}= t^{-\Gamma+1/2}\,
\nu^{-\Gamma+1},
\label{Asymptotic}
\end{equation}
with a power-law in time steeper by half a unit   
than that in frequency.
In terms of the frequently used notation,
the asymptotic behaviour satisfies,
$F_\nu(t)\propto t^{-\alpha}\,\nu^{-\beta}$
with
\begin{equation}
\alpha=\beta+1/2=p/2+1/2=\Gamma-1/2\,,
\label{indices}
\end{equation}
which is valid  for a constant density.

For a fast  falling density
beyond a distance $r_c$ seen by CBs that
encounter density bumps formed by stellar winds, or seen by CBs that 
escape
the galactic bulge or disk into the galactic halo,
$\gamma$ and $\delta$ tend to a constant
and  $r\!-\!r_c$ becomes proportional to $t-t_c$
where $r(t_c)\!=\!r_c$.
As a result, for a density profile $n\!\propto\! 1/r^2$ beyond $r_c$,
the unabsorbed synchrotron afterglow as given by
Eq.~(\ref{Fnu}) tends to,
\begin{equation}
F_\nu \propto  n^{(p+2)/4}\,\nu^{-p/2} \propto
(t-t_c)^{-(p+2)/2}\,\nu^{-p/2} =(t-t_c)^{-\Gamma}\, \nu^{-\Gamma+1}\,,
\label{Fnurm2}
\end{equation}
and satisfies the asymptotic relation,
\begin{equation}
\alpha=\beta+1=\Gamma \approx 2.1\, .
\label{indicesrm2}
\end{equation}
Thus, unattenuated optical and X-ray  AGs
of GRBs may steepen at late times to an asymptotic decay,
$\!\sim\! (t-t_c)^{-2.1}.$  
Such an achromatic steepening, which was seen in several late time
optical and X-ray AGs of Swift GRBs (see Figs.~6,7 in DD2008b), may
have been   misinterpreted as very late
`jet breaks' (e.g., Dai et al.~2008; Racusin et al.~2008a).
All the late time afterglows of Swift GRBs which are well sampled at
late time seem to satisfy 
either the asymptotic relation (\ref{indices})  or (\ref{indicesrm2}).
(see DDD2008b; DD2008b).

\section{Comparison between theory and observations}

\subsection{GRB 990123}

In Fig.~3 we compare the BATSE multipeak lightcurve of GRB 990123 in the
20-50 keV channel (Briggs et al.~1999) and its CB model description. The
count-rate in the 20-50 keV energy band was calculated from the 
integral,
$\int F_\nu\, dE/E\,,$  using Eq.~(\ref{ICSlc}) with the
best fit parameters which are listed in Table I for the 9 peaks suggested
by the multichannel BATSE data and by the BeppoSAX data (Maiorano et
al.~2005). As shown in Fig.~3 the shape of the peaks and the entire
lightcurve are well reproduced by Eq.~(\ref{ICSlc}).

In Fig.~4 we compare the early time $V$ band lightcurve of GRB 990123 as
measured by ROTSE (Akerlof et al.~1999) and the CB model expectation as
given by Eq.~(\ref{SRP}) assuming a single CB and the best fit 
parameters, $t_0=22$ s, $a=1.974$ s, $\beta_O=0.668$ and 
$A=2.49\times 10^8$ $\mu$Jy. 
The single CB approximation
was used because of the lack of information on the short time behaviour of
the optical lightcurve during the prompt emission phase. Fig.~4, however,
demonstrates that the rapid decay of the lightcurve at the end of the
prompt emission is well reproduced.

In Fig.~5 we compare the observations of the optical lightcurve of GRB 
990123 from onset (Akerlof et al.~1999) until late time ( Castro-Tirado et 
al.~1999; Galama et al.~1999; Kulkarni et al.~1999a; Fruchter et al.~1999, 
2000; Holland et al.~2000 and references therein), normalized to the 
$V$-band, and its CB model description
as given by Eq.~(\ref{Fnu}) with 
the afterglow parameters $\gamma\, \theta\!=\!0.24$, $t_0=2250$ s
and $p=1.79\,.$ Due to a 
gap in the 
data between 500 s and 15,000 s, the expected transition from a 
circumstellar density profile $\propto\! 1/r^2$ to a constant ISM density 
was not well determined, However, the gradual bending (`jet break') of the 
optical AG to an asymptotic power-law decay, $F_\nu\!\propto\! 
t^{\!-\!\beta_O\!-\!1/2} \nu^{-\beta_O}\,,$ is well reproduced with the 
expected late-time spectral index $\beta_O\!\sim\!\beta_X\!\sim\! 1.1\,.$

In Fig.~6 we compare the the lightcurve of the X-ray afterglow of GRB 
990123 in 
the 2-10 keV band, which was measured with BeppoSAX (Maiorano et al.~2005)
for $t\!<\!2.5$ days,
and its CB model description. The best fit SR lightcurve 
required 
$p\!=\!1.79$, implying $\beta_X\!=\!0.90$, consistent with 
$\beta_X\!=0.94\!\pm\!0.12$ that was inferred by Maiorano et al.~2005 
from their data.  The observed temporal power-law decay index of the 
late-time X-ray afterglow, $\alpha_X\!=\!1.46\!\pm\!0.04$ (Maiorano et 
al.~2005), obeys the CB model relation (Eq.~(\ref{indices})), 
$\alpha_X\!=\!\beta_X\!+\!1/2\!=\!1.44\!\pm\!0.13\, .$

\subsection{GRB 080319B} 

The prompt $\gamma$-ray and hard X-ray emission in GRB 080319B is composed 
of many narrow peaks (see Fig.~1 in Racusin et al.~2008), most of which 
are not well resolved, which makes the comparison between theory and 
observations for GRB 080319B less conclusive. The early-time optical 
lightcurve, was however much better sampled (Racusin et al.~2008; Wozniak 
et al.~2008) than that in GRB 990123, as shown in Fig.~7 where it is 
compared to its CB model description in terms of 3 SR peaks, each one 
described by Eq.~(\ref{SRP}) with the parameters listed in Table II. The 
decay of the prompt emission favours a single CB crossing 3 shells which 
were ejected by the progenitor star before the supernova explosion, rather 
than 3 CBs crossing a continuous pre-supernova blowing wind.

In Fig.~8 we compare the entire $R$-band (and $V$ band renormalized to the 
$R$ band) lightcurve of GRB080319 (Racusin et al.~2008)  and its CB model 
description assuming that the initially expanding 3 CBs merged into a 
single CB by the end of the prompt ICS emission of gamma-rays and hard 
X-rays around 300 s (observer time) which decelerates in roughly a 
constant density ISM. The afterglow parameters are listed in Table III. The 
`missing jet break' is hidden under the prompt emission. Shown also is the 
contribution to the $R$-band afterglow from an SN akin to SN1998bw (Galama 
et al.~1998) displaced to the GRB site.

In Fig.~9 we compare the lightcurve of 
the 0.3-10 keV X-ray afterglow of 
GRB 080319B measured with the Swift XRT (Racusin et al.~2008) and its CB 
model description, assuming a constant ISM density. The best fit 
parameters are $\gamma\,\theta\!=\!0.14 $ and $t_b<71$ s. The late time 
temporal decay of the X-ray AG is well described by a power-law with 
$\alpha_X\!=\!1.54\!\pm\! 0.04 $, except around 40,000 s, where the 
lightcurve is poorly sampled.  As expected for GRBs with large measured 
$E_p$, $E_{iso}$ and $L_p$ (DDD2008b), no AG break was observed in the XRT 
lightcurve. The wiggling of the measured lightcurve around 
a power-law decay, is probably due to variations in the 
ISM density along the CB trajectory, which we have not tried to 
parametrize.

In Fig.~10 we compare the photon spectral index of the 15-150 keV light 
curve of GRB 080319B, which was inferred from observations with the 
Swift broad alert telescope (BAT) and reported in Fig.~2 of the 
supplementary material in Racusin et al.~2008, and that expected in the CB 
model (DDD2008a). As long as the prompt hard X-ray emission is dominated 
by overlapping ICS peaks, $\Gamma\!\sim \!1$. It increases rapidly during 
the fast decay phase of the prompt emission (DDD2008a) until the 
synchrotron radiation dominates the emission and then 
$\Gamma\!\approx \!2.1$ (e.g. DDD2002). These predictions agree 
well with the observations as shown in Fig.~10.

\section{Discussion}

Accurate data that have been accumulated in recent years 
from space based and ground based observations have challenged the 
prevailing views on GRBs. This is true in particular for the brightest and 
best studied GRBs, 990123 (e.g., Maiorano et al.~2005; Corsi et al.~2005) 
and 080319B (Bloom et al.~2008; Racusin et al.~2008):  Their prompt 
multiwavelength emission cannot be explained by a single radiation 
mechanism. Their hard X-ray and $\gamma$-ray emission cannot be explained 
as SR from internal shocks generated by collisions between conical shells. 
Their prompt optical emission is not correlated with their hard X-ray and 
$\gamma$-ray emission. Their afterglows are chromatic, and roughly decay 
like a single power-law without a jet break. Their spectral and temporal 
power-law indices do not satisfy the closure relations for conical blast 
waves.

To rescue the FB model, SR as the source of the hard X-ray and 
$\gamma$-ray emission was replaced by
inverse Compton scattering of the self produced SR. 
In this so called `synchrotron self Compton' (SSC) mechanism, the prompt 
SR is produced either by internal shocks (e.g., Kumar \& Panaitescu~2008) 
or by relativistic magnetic turbulences without internal shocks (e.g. 
Kumar \& Narayan~2008). The SSC mechanism implies that the SR emission 
begins before the X-ray and $\gamma$-ray emission, and both are 
correlated. But, contrary to these expectations, the observed prompt 
optical emission lags consideraby
behind the hard X-ray and $\gamma$-ray emission and 
no temporal correlation is observed between the hard X-ray and 
$\gamma$-ray peaks and the prompt optical peaks. Moreover, the SSC model 
predicts a vanishing polarization of the prompt hard X-ray and 
$\gamma$-ray emission. The measured polarization, so far in 4 ordinary 
GRBs, suggests a large polarization (Cobb et al.~2004; Willis et 
al.~2005; Kalemci et al.~2007; McGlynn et al.~2007).

The situation concerning the FB model interpretation of the observed 
afterglow of GRBs 990123 and 080319B is similar. Although some aspects of 
the data have been explained by invoking structured jets, multiple blast 
waves propagating into the ISM, backward shocks and evolving microphysical 
parameters, no unique falsifiable predictions to test these suggestions 
were made and it is not clear whether the agreement between theory and 
some aspects of the data which was claimed is significant or results only 
from a flexible parametrization and sufficient adjustable parameters.

The situation concerning the CB model is entirely different. The 
predictions of the model were derived in fair approximations from its 
original underlying assumptions long ago. They were shown to 
predict correctly the main observed properties of GRBs and reproduce 
successfully the diverse broad band light curves of large representative 
sets of both long GRBs and XRFs (e.g., DD2004; DD2008a and references 
therein) and short hard bursts (DD2008b). In this paper, we have 
demonstrated these for GRBs 990123 and 080319B, the two brightest and so 
far the best studied gamma ray bursts, without invoking any new physics. 
They imply that GRBs 990123 and 080319B were ordinary, highly 
collimated GRBs produced in a core collapse supernova explosion within a 
high-density wind environment and observed from a very near-axis viewing 
angle. Massive wind/ejecta blown from the progenitor star before the 
supernova explosion probably created a matter-free bubble around the 
progenitor star which was filled with radiation. Inverse Compton 
scattering of this light by the jet of CBs produced the hard X-rays and 
$\gamma$-rays which preceded the broadband SR emission when the jet 
crossed the wind.  As in all other GRBS, inverse 
Compton scattering and 
synchrotron radiation, the two dominant radiation mechanisms in the 
cannonball model of GRBs, together with the burst environment -wind blown 
into a constant density ISM- provide a very simple and sufficiently 
accurate description of the multiwavelength lightcurves of their prompt 
and afterglow emissions. The dependence of the bend/break frequency on 
density variations along the jet trajectory can explain the observed 
spectral variations but obviously cannot predict it. The 
predicted  general trend 
$\beta_O(0)\!\sim\! 0.5\!\rightarrow$ $\beta_O(t)\!\sim\! 1.1$ 
at late time is well satisfiesd.

Two predictions which can be tested in future observations of extremely 
bright GRBs are as follows:  Because of their small viewing angles, the 
polarization of the prompt X-rays and $\gamma$-rays in very bright GRBs, 
such as 990123 and 080319B, is predicted to be small (Shaviv \& Dar~1995; 
DD2004), unlike that of ordinary GRBs with $\gamma\, \theta\approx 1$ 
where it is predicted to be very large.

Collisions of the CBs with the dense wind and glory, which are needed to 
produce very luminous GRBs, are expected to produce also detectable fluxes 
of high energy photons (Dado \& Dar~ 2005: Dado \& Dar, in preparation): 
Electrons from the wind and/or ISM which are 
swept into, or scattered by, the CBs should 
produce sub-GeV photons by ICS of glory light. Hadronic collisions of the 
thermal nuclei in the CBs' plasma with wind nuclei are expected to 
produce detectable fluxes of sub-TeV photons and marginal fluxes of 
neutrinos through $\pi$ production, while wind/ISM nuclei swept into or 
scattered by the CBs are expected to produce fluxes of sub-PeV 
photons which are detectable only from very nearby bursts
because of the opacity of the infrared background to high energy photons.
(Dar \& De R\'ujula~2006).

{\bf Acknowledgment:}
We thank Elisabetta Maiorano for making
available to us tabulated data of the BeppoSAX measurements
of the X-ray light curves of GRB 990123.

\section{Appendix I}

With $E_p(t)\!\approx\! E_p(0)\, t_p^2/(t_p^2\!+\!t^2)\!$,
the ICS spectral energy density satisfies,
\begin{equation}
E\, {d^2N_\gamma\over dt\,dE}(E,t) \propto e^{-E/E_p(0)}\, F(E\,t^2)\,.
\label{et2}
\end{equation}
and then, 
\begin{equation}
\int E^2\, {d^2N_\gamma\over dt\,dE}(E,t)dt \propto E^{1/2}\, 
e^{-E/E_p(0)}\,\int_0^\infty F(E\,t^2)\,d(E^{1/2}\,t)\propto 
E^{1/2}\,e^{-E/E_p(0)}\,,  
\label{tint}
\end{equation}
has a maximum at $E=E_p(0)/2=E_p(t_p)$.

\newpage
\begin{deluxetable}{llllc}
\vskip -2.cm
\tablewidth{0pt}
\tablecaption{CB model parameters of the ICS $\gamma$ peaks in GRB 990123}
\tablehead{
\colhead{Peak} & \colhead{$t_0\, [{\rm s}]$} & \colhead{$Dt\,[{\rm 
s}]$}& \colhead{$E_p\,[{\rm keV}]$} &\colhead{$A\, 
[{\rm counts\,s^{-1}}]$} 
}
\startdata
1 & -5.57  & 15.4  & 300 & $ 4.00\times 10^3$\\
2 & 19.42  & 4.28  & 1450& $ 3.95\times 10^4$\\
3 & 29.88  & 0.87  & 500 & $ 6.74\times 10^3$\\
4 & 32.95  & 5.43  & 800 & $ 2.75\times 10^4$\\
5 & 43.67  & 4.99  & 500 & $ 1.11\times 10^4$\\
6 & 52.77  & 4.30  & 450 & $ 1.05\times 10^4$\\
7 & 61.09  & 4.10  & 450 & $ 1.21\times 10^4$\\
8 & 70.35  & 5.63  & 600 & $ 1.69\times 10^4$\\
9 & 85.50  & 1.86  & 200 & $ 6.64\times 10^3$\\
\enddata
\label{t1}
\end{deluxetable}

\begin{deluxetable}{lllllc}
\vskip -5.cm
\tablewidth{0pt}
\tablecaption{CB model parameters of the early-time SR $V$-band peaks in 
GRB 080319B}
\tablehead{
\colhead{Peak} & \colhead{$t_0\, [{\rm s}]$} & \colhead{$t_{exp}\,[{\rm
s}]$} & \colhead{$A\,[\rm{Jy}]$}&
\colhead{$a\, [{\rm s}]$}& \colhead{$\beta_O$} 
}
\startdata
1 & 9.50 & 10.24   & 3755  & 3.52 & 0.70\\
2 & 32.37 & 5.83   & 878   & 3.52 & 0.70\\
3 & 39.54 & 6.96   & 1291   & 3.52 & 0.70\\
\enddata
\label{t2}
\end{deluxetable}

\begin{deluxetable}{llllc}
\vskip -5.cm
\tablewidth{0pt}
\tablecaption{CB model afterglow parameters
of GRBs 990123 and 080319B}
\tablehead{
\colhead{GRB} & \colhead{$\gamma_0\,\theta$} &
\colhead{$t_0$}& \colhead{$p$}& \colhead{$\beta_O(0)$}  }
\startdata
990123  & 0.24 & 2250 &  1.79 & 0.67\\
080319B & 0.14 &   86 &  2.15 & 0.6\\
\enddata
\label{t3}
\end{deluxetable}

\newpage
\begin{figure}[]
\centering
\epsfig{file=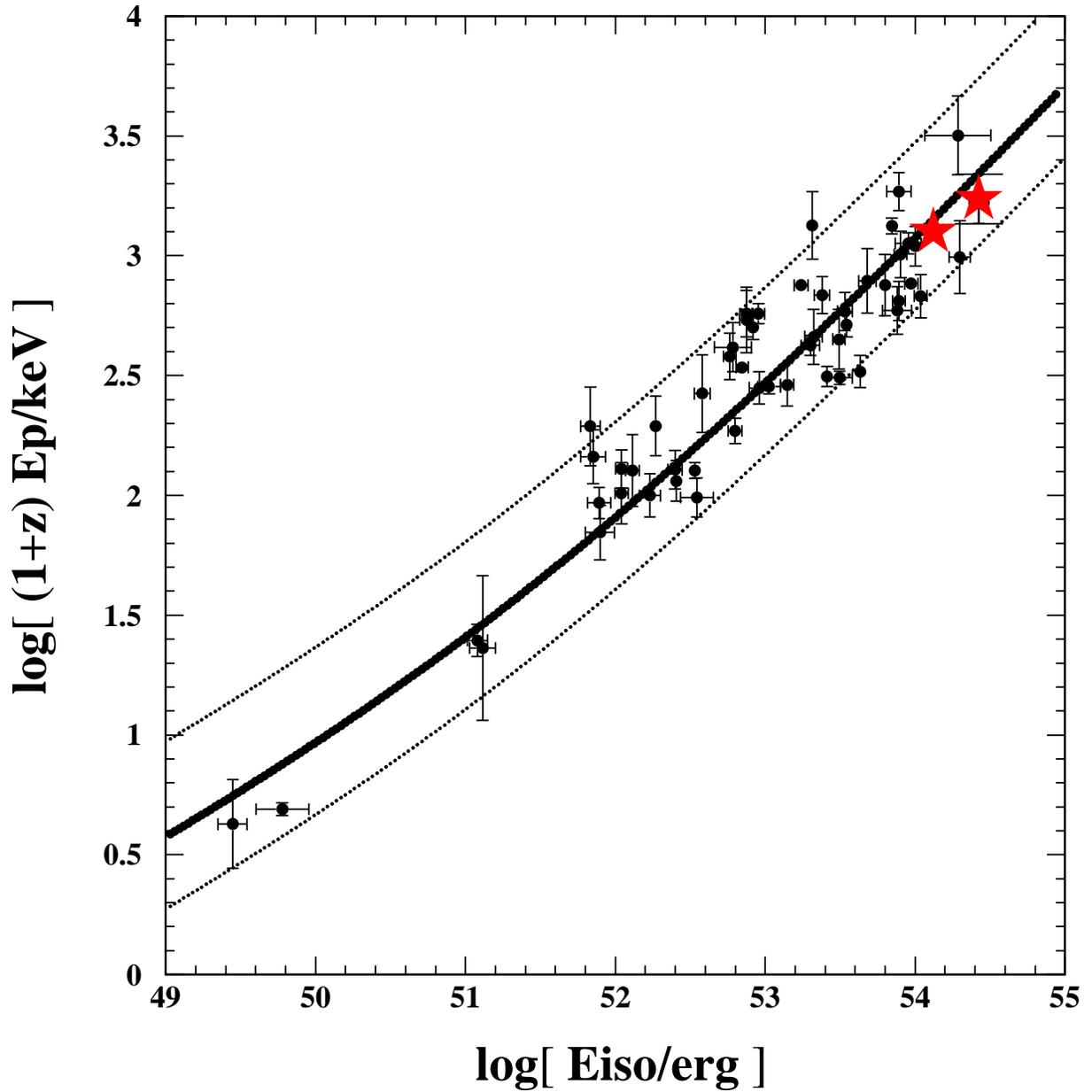,width=18cm}
\caption{The correlation between the rest frame peak photon
energy and the isotropic equivalent total gamma-ray energy of
long GRBs with known redshift. The thick line is the correlation predicted 
by the CB model (DDD2007).
GRBs 990123 and 080319B are indicated, respectively, by a large square 
and a large star}
\label{f1}
\end{figure}

\newpage
\begin{figure}[]
\centering
\epsfig{file=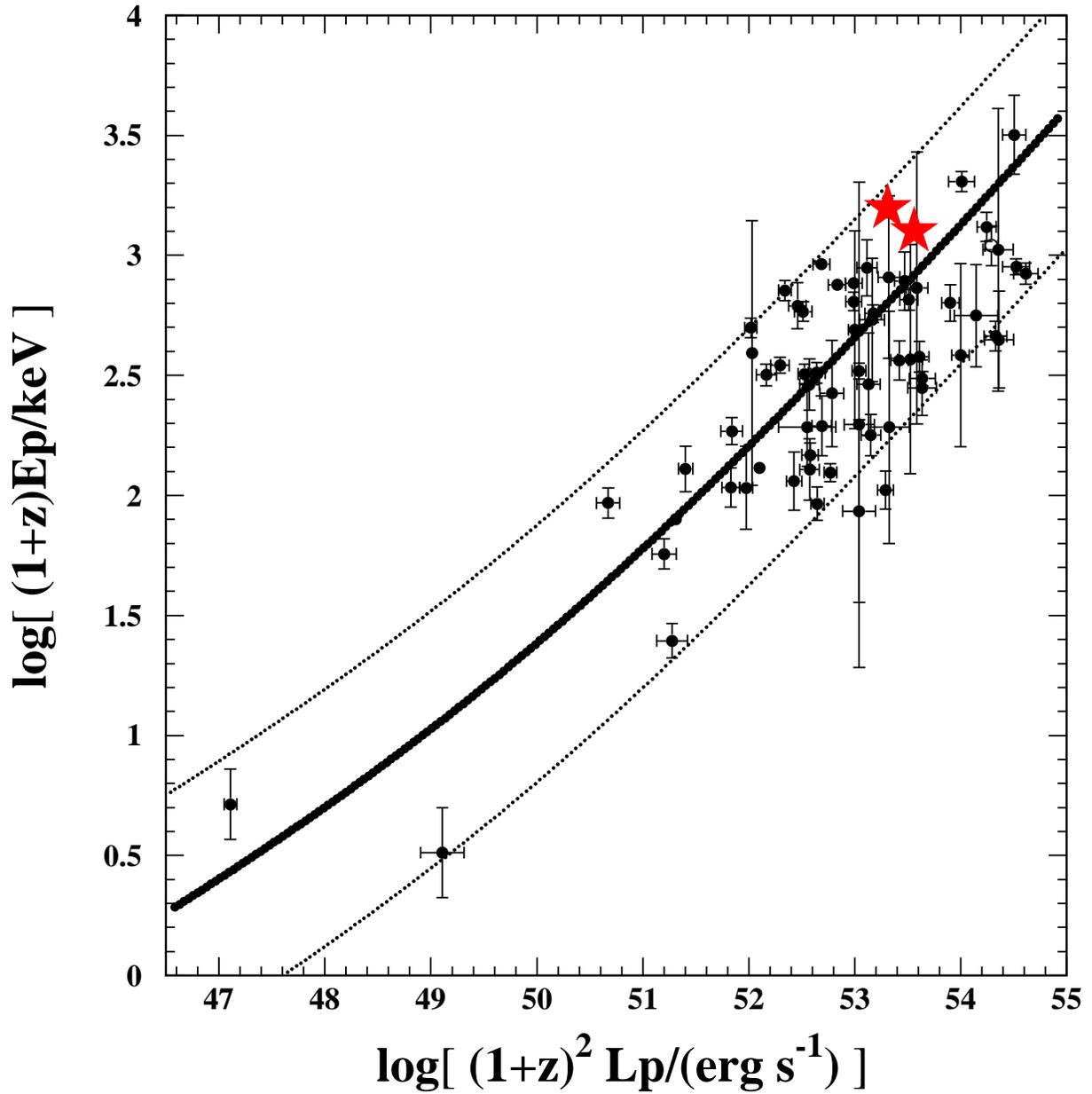,width=18cm}
\caption{The rest frame peak photon
energy plotted versus the isotropic peak  gamma-ray luminosity 
of long GRBs with known redshift.
The thick line is the correlation predicted by the CB model
(DDD2007). 
GRBs 990123 and  080319B are indicated by large red stars.} 
\label{f2}
\end{figure}

\newpage
\begin{figure}[]
\centering
\epsfig{file=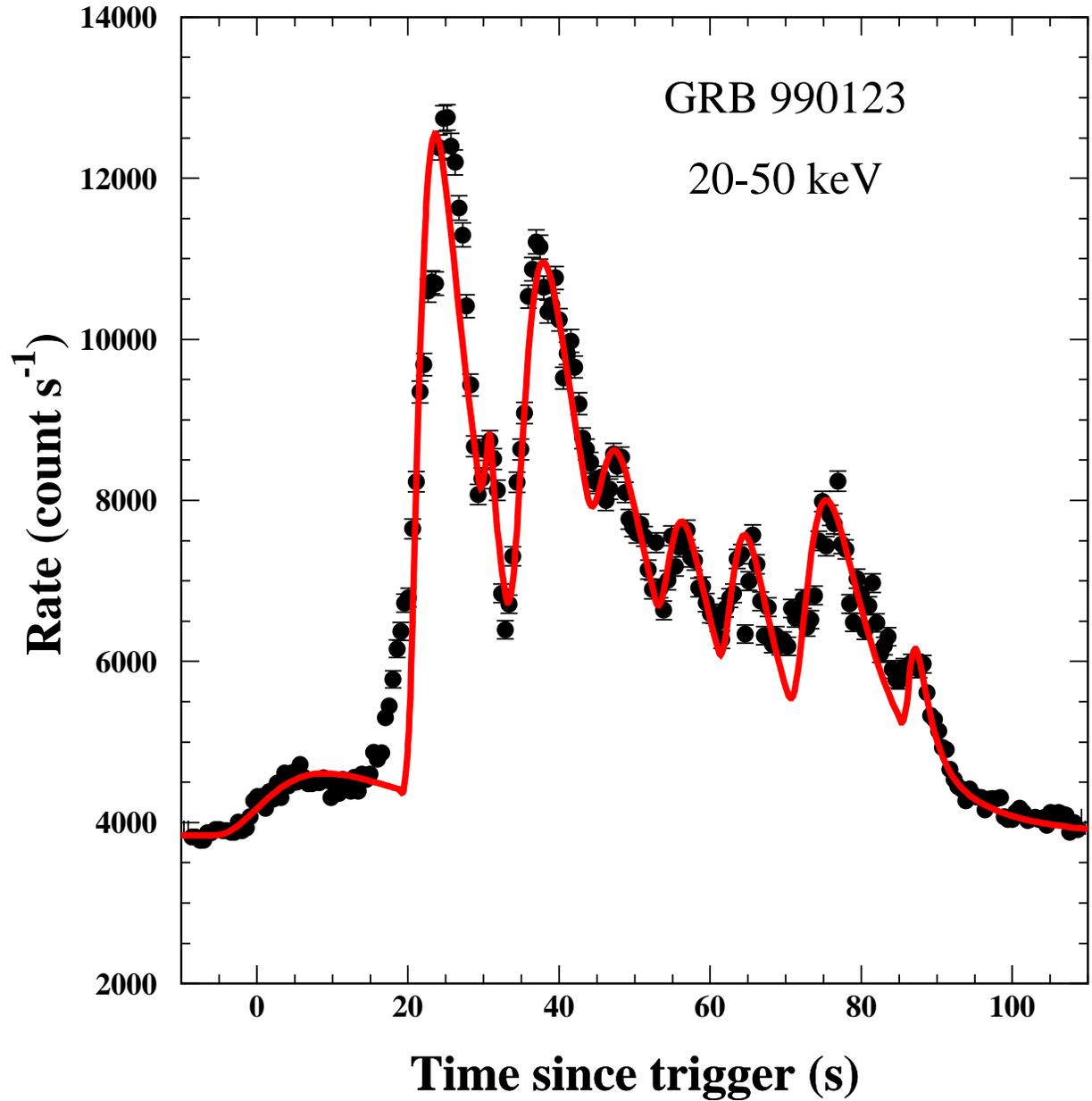,width=18cm}
\caption{Comparison between the 20-50 keV BATSE lightcurve 
of GRB990123 and its CB model description, Eq.~(\ref{ICSlc}) 
in terms of 9 ICS peaks + a constant background of 3850 counts s$^{-1}$.  
The sub-pulse superimposed on the 
decaying tail of the three major pulses may be due to 
the crossing of the 3 leading CBs through two successive wind layers 
(2 separate pre-supernova mass ejections by the progenitor star)
rather than by 3 additional CBs.}  
\label{f3}
\end{figure}

\newpage 
\begin{figure}[] 
\centering 
\epsfig{file=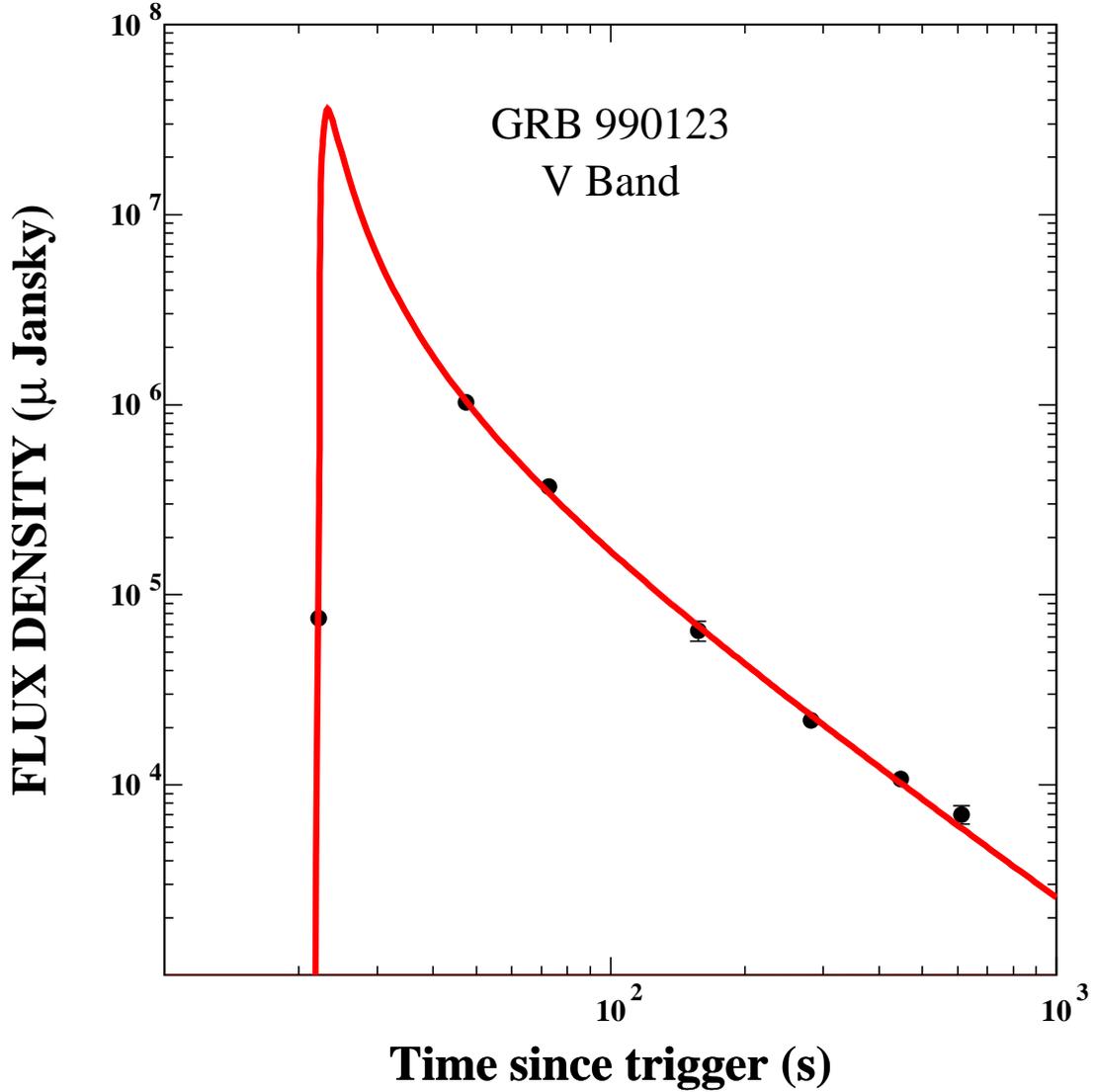,width=16cm} 
\vspace{1cm} 
\caption{Comparison 
between the ROTSE early-time $V$ band lightcurve of GRB990123 (Akerlof et 
al.~1999) and its CB model description as given by a SR flash ,
of a single CB as given by Eq.~(\ref{SRP}). 
The single CB approximation may be adequate  
because of the poor temporal sampling/temporal resolution of the optical 
lightcurve during the prompt optical emission.}
\label{f4}
\end{figure}

\newpage
\begin{figure}[]
\centering
\epsfig{file=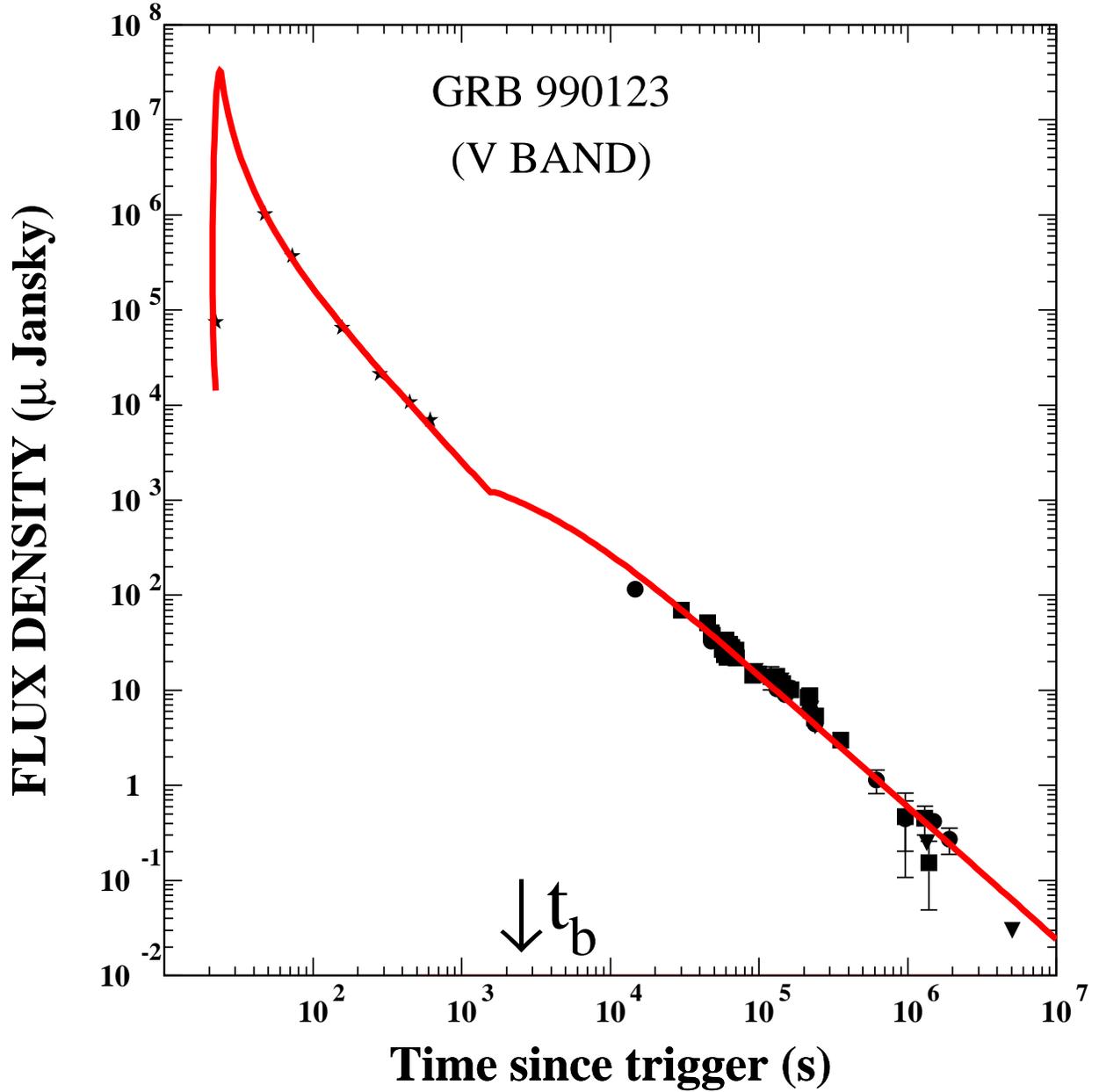,width=18cm}
\caption{Comparison between the entire $V$ band lightcurve of GRB 990123 
and its CB model description assuming a single CB moving 
in circumstellar density profile 
$\propto\! 1/r^2$  overtaken  by a constant ISM 
density around an observer time $t\!=\!1000$ s.} 
\label{f5}
\end{figure}

\newpage
\begin{figure}[]
\centering
\epsfig{file=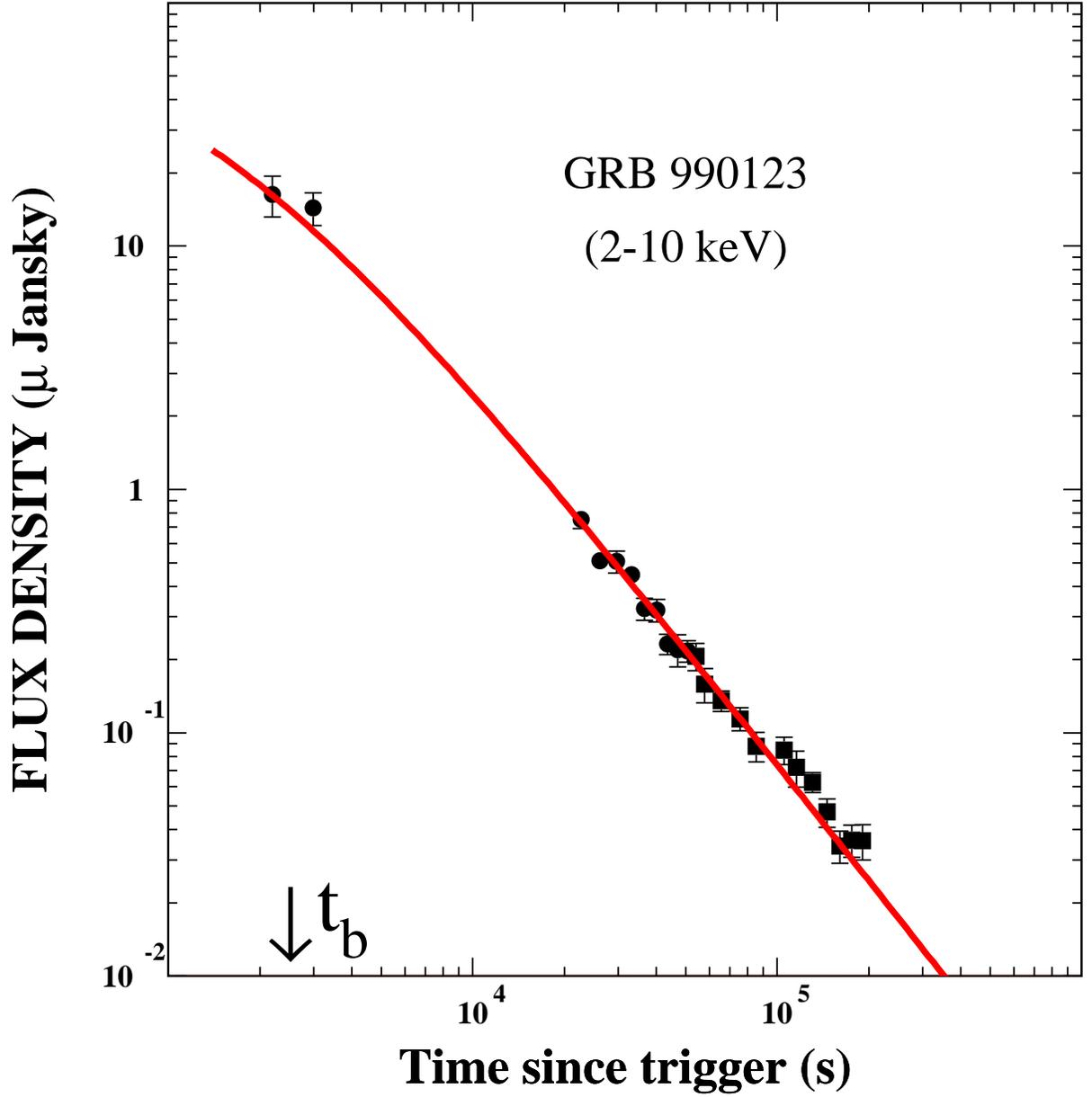,width=18cm}
\caption{Comparison between the BeppoSAX 
lightcurve of the 2-10 keV X-ray
AG of GRB990123  and its CB model description   
as given by Eq.~(\ref{Fnu}). The X-ray lightcurve 
was calculated with the same parameters used to reproduce 
the V-vand parameters.}    
\label{f6}
\end{figure}

\newpage
\begin{figure}[]
\centering
\epsfig{file=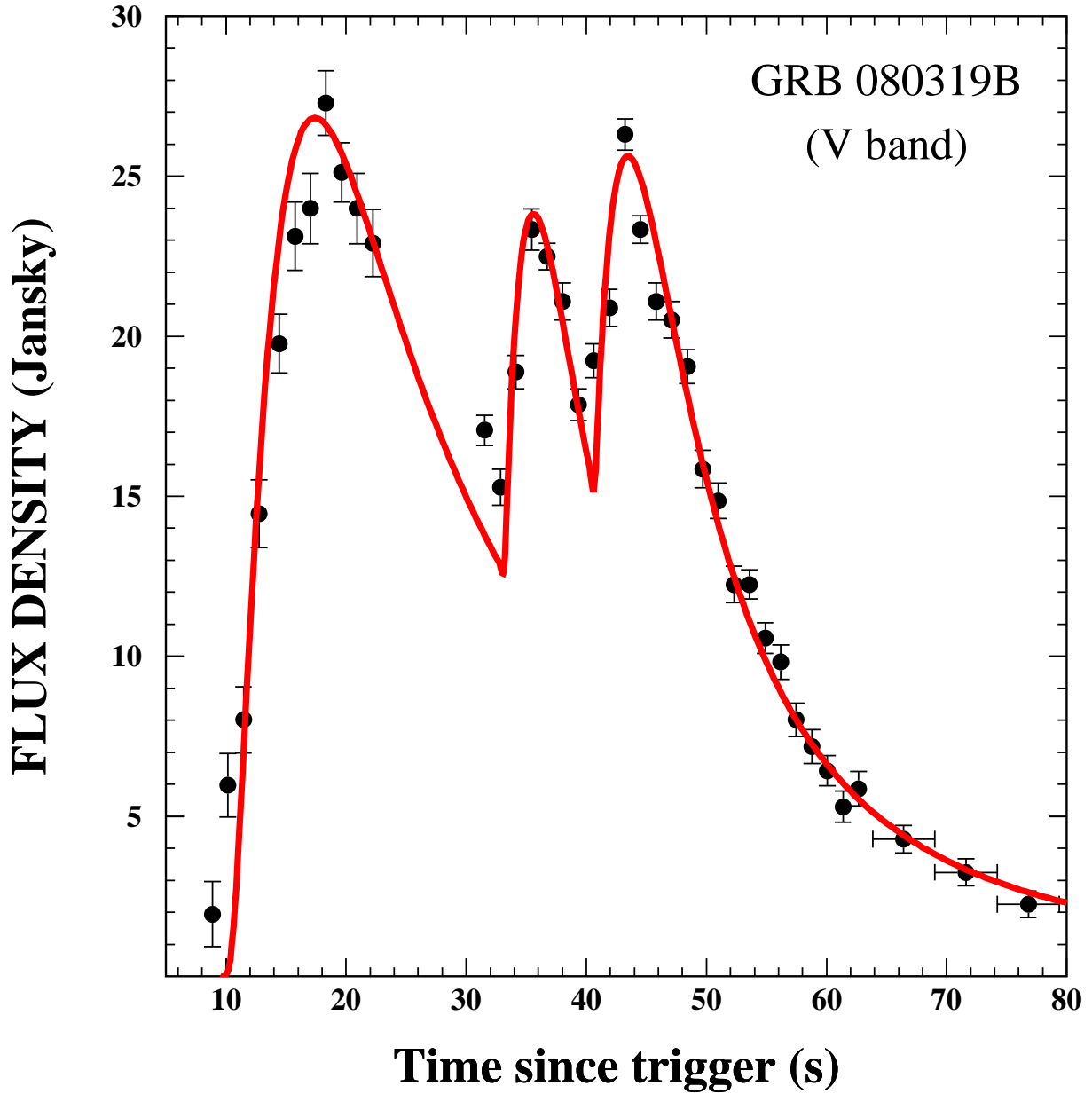,width=18cm}
\caption{Comparison between the  early time $V$-band
lightcurve of GRB080319 (Racusin et al.~2008)  and its  
CB model description in terms of 3 SR peaks. 
given by Eq.~(\ref{SRP}). The parameters used to calculate 
the CB model  lightcurve are listed in table III.
The decay of the prompt emission favours a single CB 
crossing 3 pre-supernova mass ejections by the progenitor 
star rather than 
3 CBs crossing a continuous pre-supernova blowing wind.}
\label{f7}
\end{figure}

\newpage
\begin{figure}[]
\centering
\epsfig{file=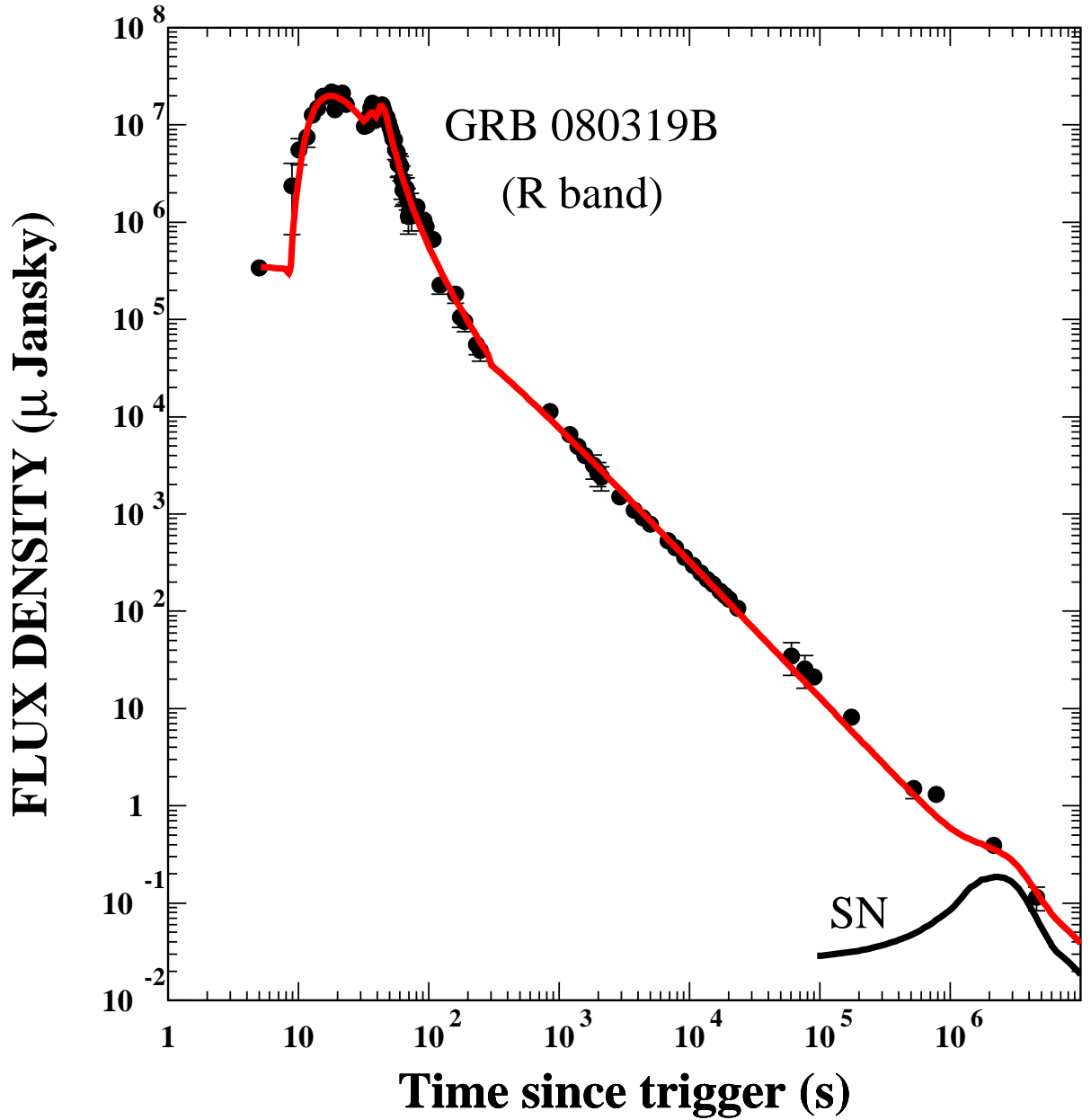,width=18cm}
\caption{Comparison between the entire $R$-band (and $V$ band renormalized 
to the $R$ band) lightcurve of GRB080319 (Racusin et al.~2008)  and its CB 
model description assuming that the initially expanding 3 CBs merge into a 
single CB moving in roughly a constant density ISM by the end of the 
prompt ICS emission of gamma-rays and hard X-rays around 300 s (observer 
time). The afterglow parameters are listed in Table IV. The afterglow 
break is hidden under the prompt emission. Shown also is the contribution 
to the $R$-band afterglow from SN akin to SN1998bw (Galama et al.~1998) 
displaced to the GRB site.}
\label{f8}
\end{figure}

\newpage
\begin{figure}[]
\centering
\epsfig{file=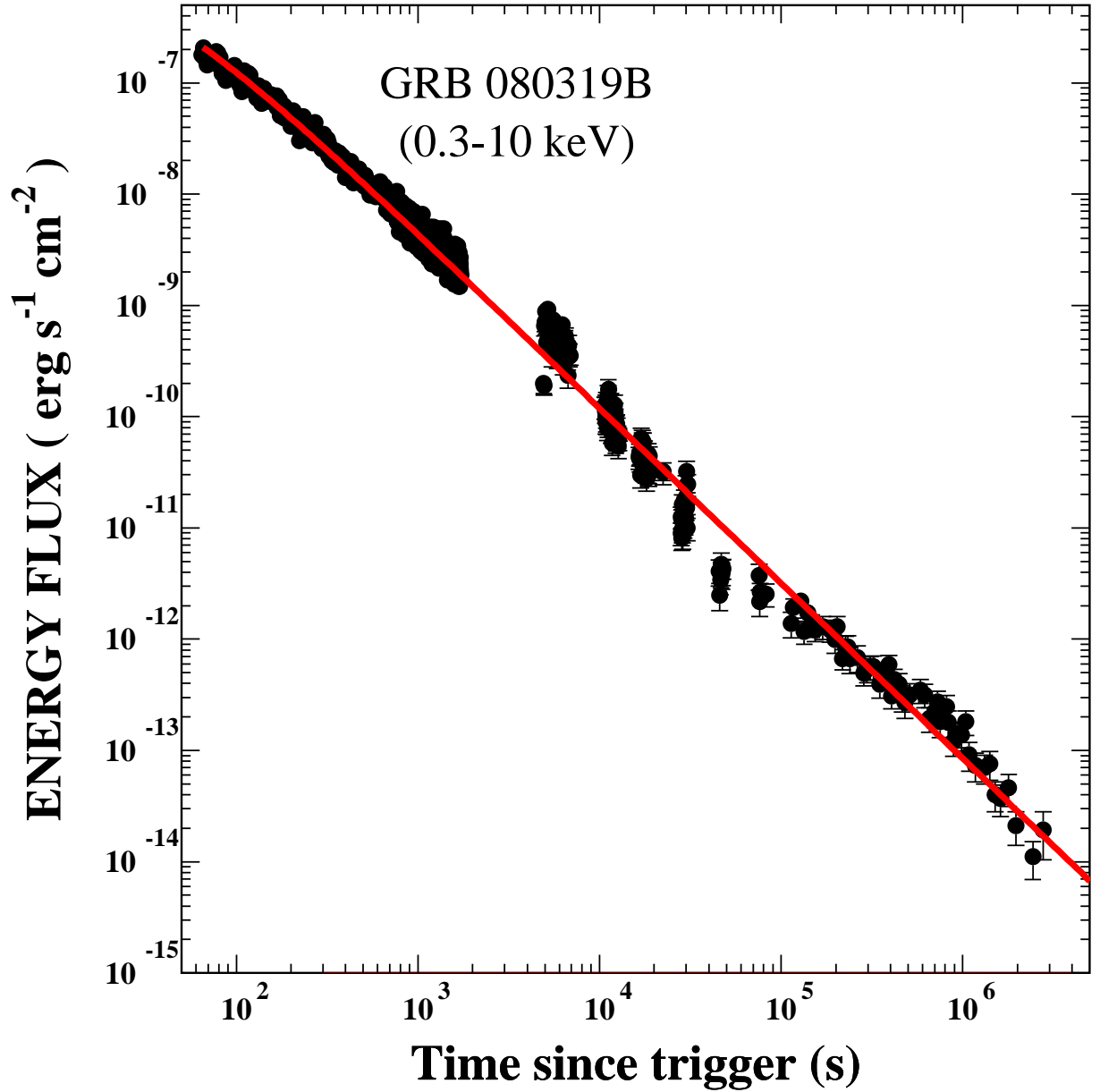,width=18cm}
\caption{
Comparison between the lightcurve
of the 0.3-10 keV X-ray afterglow of GRB 080319B measured
with the Swift X-ray telescope (XRT) and its CB model description,
Eq.~(\ref{Fnu}). For very luminous GRBs such as GRB 080319B, 
the break in the
X-ray lightcurve  is hidden under the prompt emission 
and the lightcurve is predicted to have a 
simple power-law decay.
The best fit temporal decay index, $\alpha_X=1.54\pm 0.04$,
and the observed late-time spectral index, $\beta_X=0.92\pm 0.07$,
satisfy within errors the asymptotic CB model
prediction, $\alpha_X\!=\!\beta_X+1/2$.}
\label{f9}
\end{figure}

\newpage
\begin{figure}[]
\centering
\epsfig{file=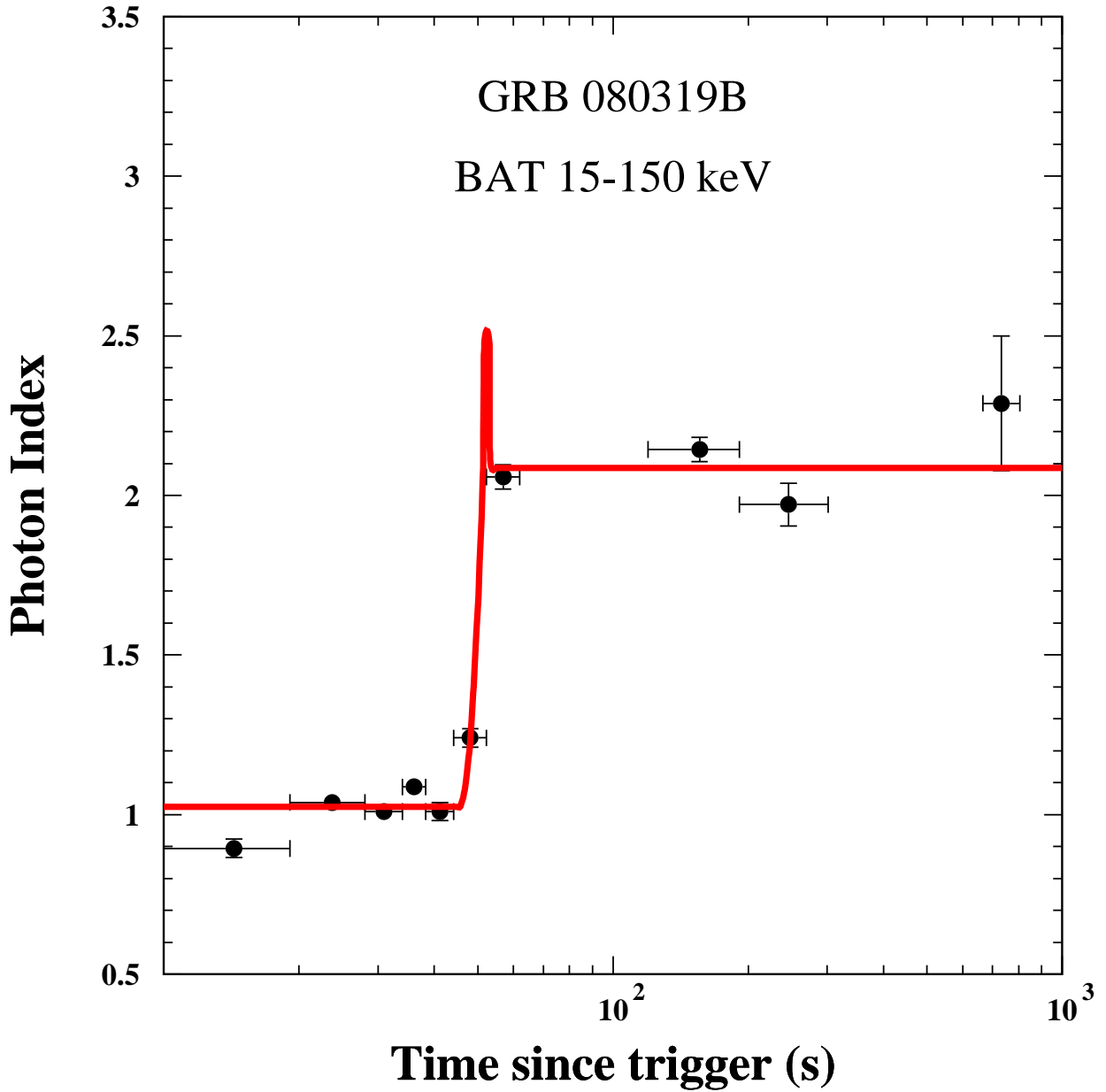,width=18cm}
\caption{Comparison between the mean photon spectral index in the 15-150 
keV band as measured with the Swift broad alert telescope (BAT) and 
reported in Fig.~ in Racusin et al.~2008, and its CB model prediction. In 
the CB model, the prompt emission is dominated by ICS of thin 
bremsstrahlung with a typical $\Gamma\!\approx \!1$, which increases 
rapidly during the fast decay phase of the prompt emission (DDD2008a) 
and becomes $\Gamma\!\approx \!2.1$, the typical value predicted by the CB 
model (e.g. DDD2002)  as soon as SR dominates the X-ray emission.}
\label{f10}
\end{figure}
\end{document}